\documentclass[journal]{IEEEtran}

\ifCLASSINFOpdf
\else
   \usepackage[dvips]{graphicx}
\fi
\usepackage{url}
\usepackage{graphicx,xcolor}
\usepackage[fleqn]{amsmath}
\usepackage{newtxtext}
\usepackage[varg]{newtxmath}
\usepackage{color}
\usepackage{algorithmic}
\usepackage{algorithm}
\usepackage{comment}

\renewcommand{\Vec}[1]{\textrm{\boldmath $#1$}} 
\def\pt#1{\left(#1\right)} 

\usepackage{graphicx}

\begin{document}

\title{Incremental Text-to-Speech Synthesis Using Pseudo Lookahead with Large Pretrained Language Model}

\author{Takaaki Saeki, \IEEEmembership{Student Member, IEEE}, Shinnosuke Takamichi, and Hiroshi Saruwatari, \IEEEmembership{Member, IEEE}
\thanks{
The authors are with
Graduate School of Information Science and Technology,
The University of Tokyo,
7-3-1 Hongo, Bunkyo-ku, Tokyo 113-8656, Japan
(e-mail: \{ takaaki\_saeki, shinnosuke\_takamichi, hiroshi\_saruwatari \}@ipc.i.u-tokyo.ac.jp).
}
\thanks{
Part of this work was supported by JSPS KAKENHI Grant Number 17H06101 and 19H01116, and the MIC/SCOPE \#182103104.
}
}

\markboth{Accepted for IEEE Signal Processing Letters}
{Shell \MakeLowercase{\textit{et al.}}: Bare Demo of IEEEtran.cls for IEEE Journals}
\maketitle

\begin{abstract}

This letter presents an incremental text-to-speech (TTS) method that performs synthesis in small linguistic units while maintaining the naturalness of output speech.
Incremental TTS is generally subject to a trade-off between latency and synthetic speech quality.
It is challenging to produce high-quality speech with a low-latency setup that does not make much use of an unobserved future sentence (hereafter, ``lookahead''). 
To resolve this issue, we propose an incremental TTS method that uses a pseudo lookahead generated with a language model to take the future contextual information into account without increasing latency.
Our method can be regarded as imitating a human's incremental reading and uses pretrained GPT2, which accounts for the large-scale linguistic knowledge, for the lookahead generation.
Evaluation results show that our method 1) achieves higher speech quality than the method taking only observed information into account and 2) achieves a speech quality equivalent to waiting for the future context observation.

\end{abstract}


\IEEEpeerreviewmaketitle
\vspace{-1mm}
\section{Introduction}
Simultaneous speech-to-speech translation (SST)~\cite{bangalore-etal-2012-real,sudoh2020simultaneous} enables interactive speech communication without language barriers.
It consists of three modules that perform incremental processing: automatic speech recognition (ASR), machine translation (MT), and text-to-speech synthesis (TTS).
Recent advances in deep learning have made remarkable progress in the quality of TTS, as well as in ASR and MT.
It is now possible to artificially generate high-quality speech comparable to human natural speech by modeling time-series information in the whole sentence with deep neural networks.
In contrast to the typical sentence-level TTS frameworks, incremental TTS requires handling small linguistic segments at the level of a few words, which makes it more challenging.
Therefore, incremental TTS suffers from a trade-off between the naturalness of the output speech and latency in the synthesis.
Low-latency incremental TTS should process the current segment using only an observed segment, rather than waiting for an unobserved future segment ahead of the current segment (hereafter, ``lookahead'').
However, this makes it difficult to output a speech segment that leads naturally to the lookahead, causing synthetic speech quality to deteriorate.

This letter proposes a method to perform high-quality and low-latency synthesis using a pseudo lookahead generated with a large-scale pretrained language model.
When we humans need to read an unknown sentence sequentially (e.g., reading a new book), we can predict future information on the basis of the observed segment. 
Then we can read out the segment so that it is naturally connected to the past observed and predicted contexts.
To computationally imitate this mechanism, our method predicts the lookahead using pretrained GPT2~\cite{radford2019language}, which is trained on datasets from various domains.
It can enhance synthetic speech quality without increasing the latency by using the pseudo lookahead as the future contextual information instead of waiting for the ground-truth lookahead.
This method is effective and applicable to other incremental TTS frameworks (e.g., prefix-to-prefix decoding~\cite{Ma20prefix}).
The model architecture is a Tacotron2~\cite{shen17tacotron2}-based end-to-end TTS model, which incorporates a contextual embedding network~\cite{ppspeech} that considers the past observed and the future unobserved contexts, and consistently trains the entire model to achieve the high-quality synthesis of the current segment.
We also propose a language model-guided fine-tuning method to estimate the contextual embedding that is more suitable for the predicted sentence with GPT2.
Evaluation results show that our method 1) achieves higher speech quality than the method taking only observed information into account and 2) achieves a speech quality equivalent to waiting for the future context observation.

\vspace{-1mm}
\section{Related works}
In recent years, the quality of TTS has dramatically improved with the shift from cascade statistical parametric speech synthesis~\cite{tokuda02englishhts,zen09,zen13dnn} to end-to-end TTS~\cite{wang17tacotron,shen17tacotron2,Nihan19transformer}, which generates a mel-spectrogram from text with a single model.
Several studies have focused on incremental TTS with end-to-end architectures~\cite{Yanagita19itts,Ma20prefix,Stephenson2020,Mohan2020IncrementalTT}.
The first attempt at end-to-end neural incremental TTS~\cite{Yanagita19itts} uses a Tacotron~\cite{wang17tacotron}-based model to achieve high-quality synthesis.
This method is a segment-level incremental TTS just like ours, and it has difficulty generating natural speech segments because the synthesis process is isolated from the past observed and unobserved future contexts, as we discuss in Section~\ref{sec:evaluation}.
Ma~et~al. proposed a prefix-to-prefix framework for incremental TTS with a lookahead-$k$ policy that waits to observe future $k$ words and synthesizes a current segment~\cite{Ma20prefix}.
In contrast to this framework, our work focuses on instantly synthesizing speech from a current segment without waiting for the lookahead.
The TTS model we use has a contextual embedding network designed in the prior work for sentence-level TTS~\cite{ppspeech}.
This method aims at parallel synthesis by focusing on intonational phrases, and both pre- and post-phrases of an input phrase can be used for the inference process, while we can only use the pre-segment of the current segment.

\begin{figure}[t]
  \centering
  \includegraphics[width=0.90\linewidth, clip]{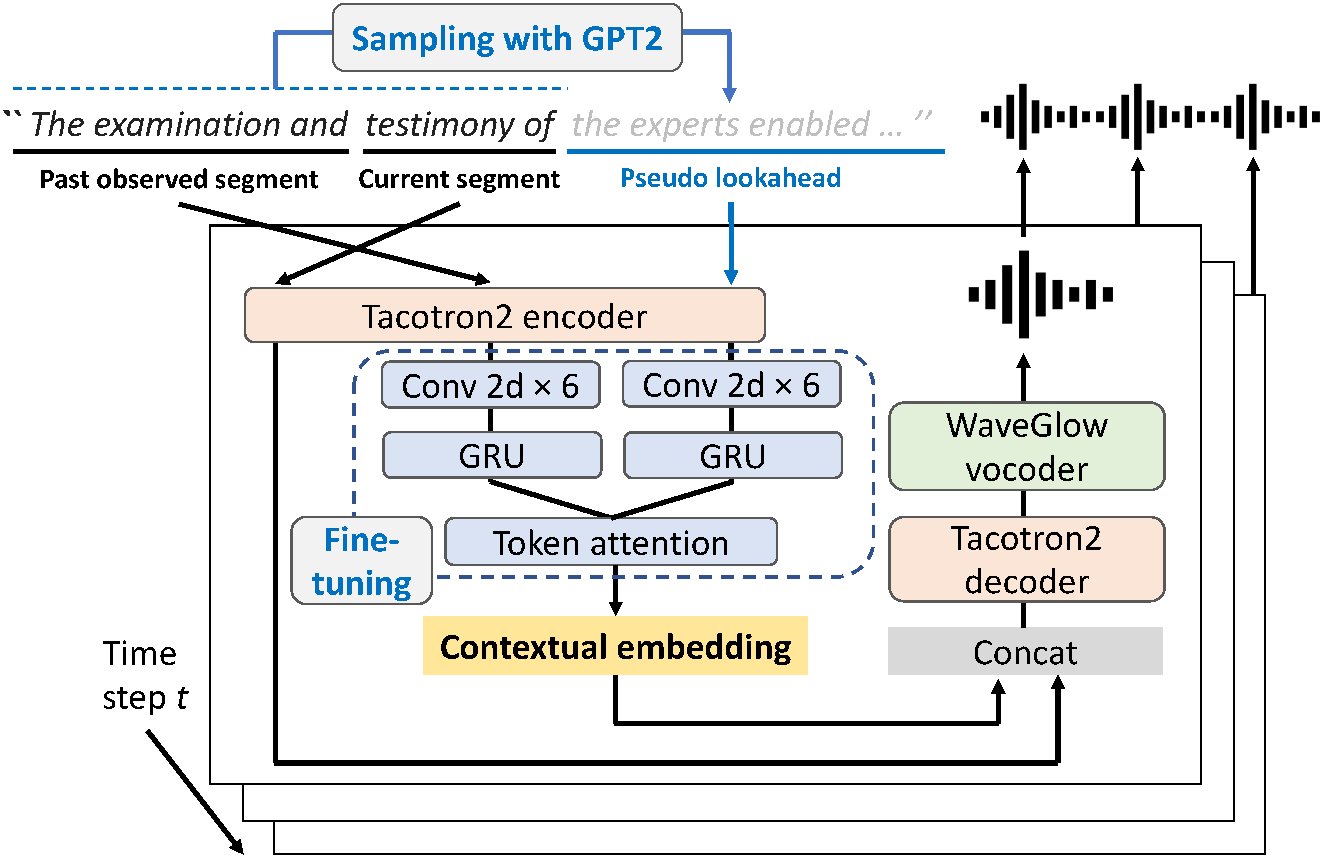}
  \caption{Model architecture of proposed method with contextual embedding network to consider past observed segment and pseudo lookahead.}
  \label{fig:model}
  \vspace{-8pt}
\end{figure}

\vspace{-1mm}
\section{Method}\label{methods}
This section describes our proposed incremental TTS method.
Section~\ref{methods_predict} presents an inference algorithm, which integrates a sentence generation with GPT2.
Section~\ref{methods_model} describes a model architecture for generating a speech segment considering both past observed and future unobserved contexts. Section~\ref{methods_finetune} presents a language model-guided fine-tuning method.
Finally, Section~\ref{methods_discussion} provides a detailed analysis of the proposed method.

\vspace{-1mm}
\subsection{Incremental synthesis with pseudo lookahead}\label{methods_predict}
First, we define the synthetic unit for incremental TTS as ``the current segment'', which consists of $N$ words.
In the time step $t$, $\Vec{w}_{1:Nt} = \Vec{w}_{1}, \cdots, \Vec{w}_{n}, \cdots, \Vec{w}_{Nt}$ represents ``the observed segment'' and the last $N$-word sequence $\Vec{w}_{N(t-1)+1:Nt} = \Vec{w}_{N(t-1)+1}, \cdots, \Vec{w}_{Nt}$ is a current segment, where $\Vec{w}_{n}$ denotes the $n$-th word.
Furthermore, we define $\Vec{w}_{1:N(t-1)} = \Vec{w}_{1}, \cdots, \Vec{w}_{N(t-1)}$ as ``the past observed segment'' to distinguish between the observed segments with and without the current segment.
GPT2~\cite{radford2019language} is an auto-regressive language model that assumes the probability distribution of an $M$-word sequence $\Vec{w}_{1:M}$ can be decomposed into the product of conditional probabilities, as
\begin{align}
    p(\Vec{w}_{1:M}) = {\displaystyle \prod_{m=1}^{M}}p\pt{\Vec{w}_{m}|\Vec{w}_{1:m-1}}.
\end{align}
In accordance with this modeling, we can obtain a future $L$-word sequence $\hat{\Vec{w}}_{Nt+1:Nt+L} = \hat{\Vec{w}}_{Nt+1}, \cdots, \hat{\Vec{w}}_{Nt+L}$ by sampling from the probability distribution $p(\Vec{w}_{Nt+1:Nt+L} | \Vec{w}_{1:Nt})$, where $\hat{\Vec{w}}_{Nt+1:Nt+L}$ becomes the ``pseudo lookahead'' used for the future contextual information of incremental TTS.
Since the TTS model uses a character or phoneme sequence instead of the word sequence $\Vec{w}_{n}$, we define the character or phoneme sequence corresponding to $\Vec{w}_{n}$ as $\Vec{x}_{n}$.
Defining the TTS model as $G(\cdot)$, the output mel-spectrogram $\Vec{y}_{t}$ can be obtained by
\begin{align}
    \Vec{y}_{t} = G\pt{\Vec{x}_{N(t-1)+1:Nt} | \Vec{x}_{1:N(t-1)}, \hat{\Vec{x}}_{Nt+1:Nt+L}, \Vec{\theta}_{G}},
\end{align}
where $\Vec{\theta}_{G}$ denotes parameters of $G(\cdot)$.
When we define $\Vec{z}_{t}$ as the waveform synthesized from mel-spectrogram $\Vec{y}_{t}$, waveform synthesis is performed using WaveGlow~\cite{prenger2018waveglow} vocoder $V(\cdot)$ as:
\begin{align}
    \Vec{z}_{t} = V\pt{\Vec{y}_{t} | \Vec{\theta}_{V}},
\end{align}
where $\Vec{\theta}_{V}$ denotes parameters of $V(\cdot)$.
We incrementally synthesize the output speech by concatenating $\Vec{z}_{t}$ to the audio waveform $\Vec{z}_{1:t-1}$ that has been output so far.

The naturalness of synthesized speech and latency in synthesis highly depend on $N$.
Although it is desirable to make $N$ smaller to develop low-latency incremental TTS, the naturalness of output speech degrades as $N$ decreases.
In our preliminary experiment, we found that our proposed method often outputs unintelligible speech with $N=1$, as reported in Yanagita et al.'s work~\cite{Yanagita19itts}.
Therefore, we set $N$ to two for our evaluation in this study.

\vspace{-1mm}
\subsection{TTS model architecture}\label{methods_model}
The incremental TTS model we use is a Tacotron2~\cite{shen17tacotron2}-based end-to-end model conditioned on both past observed and unobserved future segments.
It has a module for contextual embedding~\cite{ppspeech}, as shown in Fig.~\ref{fig:model}.
Character or phoneme sequences of the current segment $\Vec{x}_{N(t-1)+1:Nt}$, the past observed segment $\Vec{x}_{1:N(t-1)}$ and the lookahead $\hat{\Vec{x}}_{Nt+1:Nt+L}$ are sent to the Tacotron2 encoder. 
When we define $\Vec{h}_{1:N(t-1)}$ and $\hat{\Vec{h}}_{Nt+1:Nt+L}$ as hidden states of the past observed segment and the pseudo lookahead, respectively, $\Vec{h}_{1:N(t-1)}$ and $\hat{\Vec{h}}_{Nt+1:Nt+L}$ are separately sent to additional encoders, which stack six 2-D convolutional layers and a gated recurrent unit (GRU) layer.
The outputs are concatenated and sent to a token attention layer based on a global style token~\cite{wang2018style}.
We define the network that estimates contextual embedding from the output of the Tacotron2 encoder as the ``contextual embedding network'' and denote it as $F(\cdot)$.
We can obtain the contextual embedding with the pseudo lookahead $\Vec{e}_{\mathrm{pseudo}}$ as:
\begin{align}
    \Vec{e}_{\mathrm{pseudo}} = F(\Vec{h}_{1:N(t-1)}, \hat{\Vec{h}}_{Nt+1:Nt+L}).
\end{align}
Then $\Vec{e}_{\mathrm{pseudo}}$ is replicated and concatenated with every state of $\Vec{h}_{N(t-1)+1:Nt}$ to form the input of the decoder, where $\Vec{h}_{N(t-1)+1:Nt}$ is the hidden state of the current segment.
The contextual encoders for the past observed and unobserved future segments share the same parameters, and we used the same values for the hyperparameters of the contextual embedding network as Cong et al.~\cite{ppspeech}.
By jointly training the contextual embedding network and the encoder-decoder network of Tacotron2,  we attain natural speech segments by considering both the past and future contexts.

When we train the TTS model, we use the ground-truth sentence in the training data as the unobserved future segment.
To extract the past observed segment $\Vec{x}_{\le N(t-1)}$, the current segment $\Vec{x}_{N(t-1)+1:Nt}$, and the unobserved future segment $\Vec{x}_{Nt+1 \ge}$, we use the sliding text window~\cite{ppspeech} with a fixed number of words, whereas the original one uses the text window with a fixed number of phrases.
Finally, we extract the ground-truth waveform corresponding to the current segment with forced alignment.

\vspace{-1mm}
\subsection{Language model-guided fine-tuning}\label{methods_finetune}
As we described in Section~\ref{methods_predict}, the lookahead prediction makes use of linguistic knowledge of a large pretrained language model for incremental TTS.
This method, however, results in a mismatch between the ground-truth lookahead used during training and the pseudo lookahead during inference.
In other words, the TTS model cannot fully utilize the pseudo lookahead generated with GPT2 since the TTS model does not take the lookahead prediction into account.

Therefore, we propose a language model-guided fine-tuning method to use the pseudo lookahead for incremental TTS more effectively.
In contrast to the training procedure described in Section~\ref{methods_model}, we use the pseudo lookahead generated with GPT2 as the unobserved future segment during the fine-tuning.
GPT2 generates the unobserved future segments as training data by using the past observed segments and the current segments extracted with the sliding text window.
Let $\Vec{e}_{\mathrm{pseudo}}$ be the contextual embedding obtained by using the pseudo lookahead, and $\Vec{e}_{\mathrm{truth}}$ be the contextual embedding with the ground-truth lookahead.
Our goal is to enable the contextual embedding network to use the pseudo lookahead for the contextual information to the same extent as the actual lookahead.
Therefore, we add the additional loss $L_{\mathrm{sim}}$ to the loss for the TTS model training with a weight $\alpha_{\mathrm{sim}}$ to maximize the cosine similarity between $\Vec{e}_{\mathrm{pseudo}}$ and $\Vec{e}_{\mathrm{truth}}$, as
\begin{align}
    \alpha_{\mathrm{sim}} \cdot L_{\mathrm{sim}} = \alpha_{\mathrm{sim}} \cdot \pt{1 - Sim\pt{\Vec{e}_{\mathrm{pseudo}}, \Vec{e}_{\mathrm{truth}}}},
\end{align}
where $Sim(\cdot)$ denotes the cosine similarity.
Then, unlike the TTS model training, we fix the weights of both the encoder and decoder networks of Tacotron2, and we only train the contextual embedding network.
These operations help the TTS model to consider the contextual information in a way that better fits the prediction of GPT2.

\vspace{-1mm}
\subsection{Discussion}\label{methods_discussion}

\begin{figure}[t]
  \centering
  \includegraphics[width=0.90\linewidth, clip]{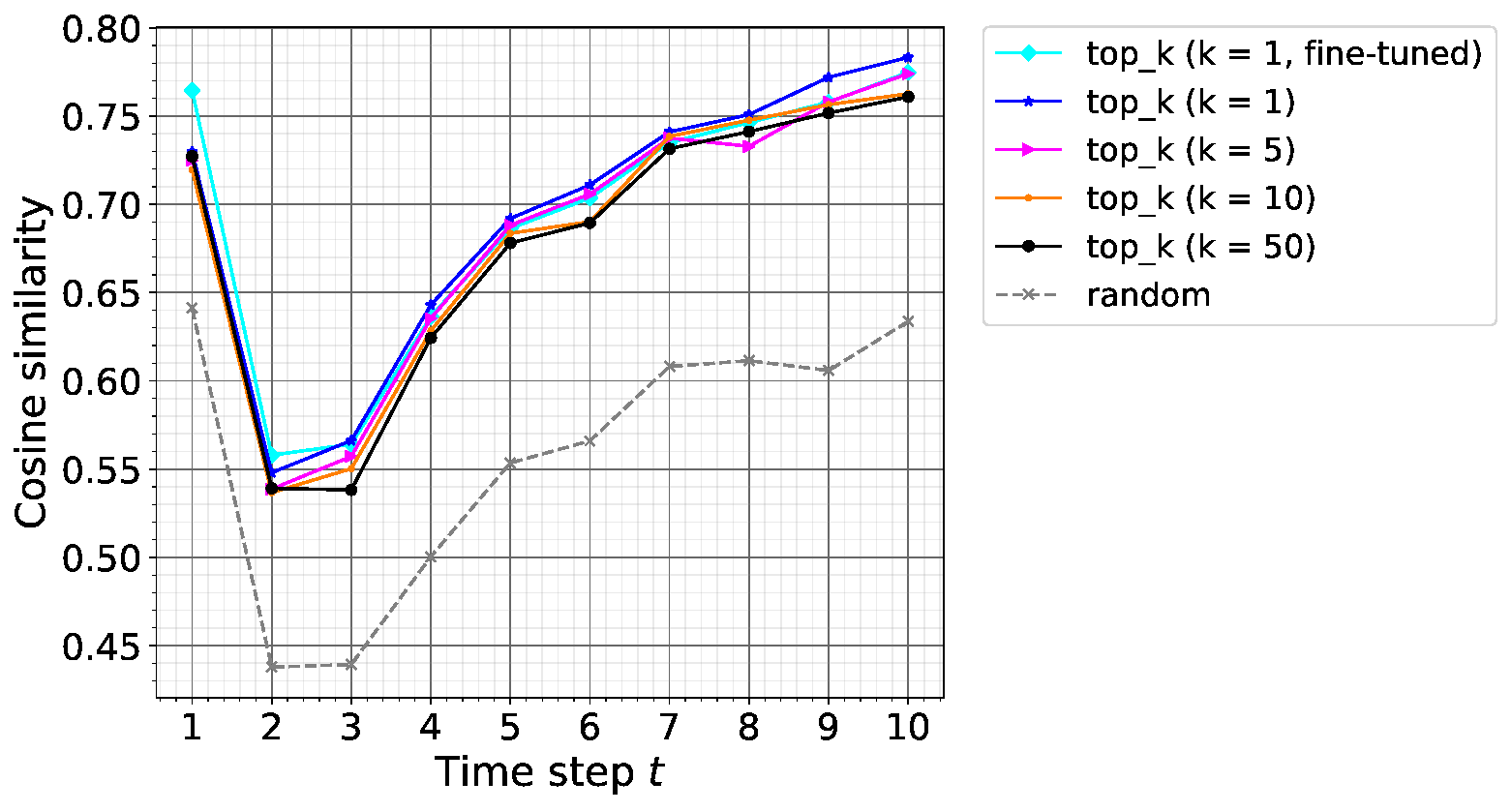}
  \caption{Average cosine similarity for time step $t$. We 1) investigate the effect of $k$ in the top-$k$ sampling and 2) compare the cases with and without the proposed fine-tuning method for $k=1$.}
  \label{fig:cossim}
  \vspace{-8pt}
\end{figure}

First, we analyze how close the pseudo lookahead generated with GPT2 is to the ground-truth lookahead.
For each time step $t$, we calculate the average cosine similarity between the contextual embedding obtained with the pseudo lookahead and that with the ground-truth lookahead.
When the cosine similarity is high, the pseudo lookahead is expected to produce an equivalent effect on the synthesized speech to the actual observation of the ground-truth lookahead.
Furthermore, we investigate the effect of the sampling strategy of GPT2.
GPT2 generates a sentence by randomly sampling from the distribution of the most probable $k$ words, which is called top-$k$ sampling~\cite{fan2018topk}.
When we set a large value to $k$, GPT2 performs random sampling from various word candidates.
When $k$ is one, GPT2 uses deterministic generation on the basis of the maximum likelihood.

Fig.~\ref{fig:cossim} shows the analysis results.
Note that we used the same experimental conditions as those described in Section~\ref{sec:evaluation_condition}. 
The label ``top\_$k$ ($k=K$)'' ($K=1, 5, 10, 50$) denotes the case where top-$k$ sampling with $k=K$ is used without the fine-tuning method, and ``top\_$k$ ($k=1$, fine-tuned)'' represents the case where top-$k$ sampling with $k=1$ is applied with the fine-tuning method.
The label ``random'' denotes the case without a language model, where we used random English words as the lookahead sentences.
Comparing the results with ``top\_$k$ ($k=K$)'' and ``random'', we can see that the lookahead generation with all $k$ cases led to better scores than the ``random'' case, demonstrating the effectiveness of the pseudo lookahead with GPT2.
Furthermore, we found that the contextual embedding obtained with the pseudo lookahead tended to become closer to the ground-truth, as the value of $k$ decreased.
We also conducted subjective evaluations on synthetic speech quality for this aspect and found that the $k=1$ case produced output speech with significantly higher naturalness than the $k=5$ and $k=50$ cases.
Intuitively, a large value of $k$ enables diverse sentence generation, and a small $k$ produces objectively plausible sentences.
The results suggest that we need to make the value of $k$ small for incremental TTS on a regular speech corpus.
Note that the cosine similarity for $t=1$ was higher than that for all the $t=2-5$ cases even in ``random''.
This result indicates that, when $t=1$, the contextual embedding network focused much more on ``the beginning of the sentence'' than the future unobserved content.
Examining ``top\_$k$ ($k=1$, fine-tuned)'', the cosine similarity with the fine-tuning was better for $t=1$ and $2$ and became lower than that in some non-fine-tuning cases as $t$ increased.
Since the fine-tuning method takes into account the pseudo lookahead with GPT2 during training, it could estimate the contextual embedding more closely to that with the ground-truth lookahead when the input segments were not well observed, i.e., at the beginning of the sentence.
However, as $t$ increased and the segments of the original sentence came in, the cosine similarity with the fine-tuning converged to the same level as that without the fine-tuning.

\vspace{-2mm}
\section{Experimental evaluations}\label{sec:evaluation}

\subsection{Experimental conditions}\label{sec:evaluation_condition}
We used LJSpeech~\cite{ljspeech17}, a dataset consisting of 13,100 short audio clips of a female English speaker lasting approximately 24~hours.
We randomly selected 100 and 500~sentences from the entire dataset for validation and test sets, respectively, and used the rest as a training set.
When extracting a mel-spectrogram from each audio clip with short-time Fourier transform, we used 1024-sample frame size, 256-sample hop size, a Hann window function, and an 80~channel mel-filterbank at a sampling frequency of 22.05~kHz. 
To use contextual information in the training process, we used the sliding text window described in Section~\ref{methods_model} with the window length $3$ and the hop size $1$.
As described in Section~\ref{methods_predict}, we set the number of words in each input segment $N$ to two in the synthesis process.
When extracting a waveform of each current segment, we used a Kaldi-based forced-alignment toolkit~\cite{gentle17}.
We used the pretrained GPT2\footnote{\url{https://github.com/graykode/gpt-2-Pytorch}} and WaveGlow\footnote{\url{https://github.com/NVIDIA/waveglow}} models for the evaluation.
On the basis of a preliminary experiment in which we calculated the cosine similarity values over time steps for different $L$, we selected $L=5$ as the setting that produces the closest lookahead to using the ground-truth future segment.
When performing the sampling operation with GPT2, we applied top-$k$ sampling with $k=1$ in all cases.
We trained the TTS model with a batch size of $160$ distributed across four NVIDIA V100 GPUs for 76000~iterations, for which we observed the convergence in all the training cases.
When performing the fine-tuning, we trained only the contextual embedding network with a batch size of $32$ on a NVIDIA Geforce GTX 1080Ti GPU for 4000~iterations, where we used $\alpha_{\mathrm{sim}} = 10^{-3}$.
We used the Adam~\cite{kingma14adam} optimizer with $\beta_{1} = 0.9$, $\beta_{2}=0.999$, and $\epsilon = 10^{-6}$.
We set a learning rate of $10^{-3}$ and $10^{-4}$ in the TTS model training and the fine-tuning, respectively, applying $L_{2}$ regularization with weight $10^{-6}$.

\vspace{-2mm}
\subsection{Evaluation cases}\label{sec:evaluation_cases}
To investigate the effectiveness of lookahead prediction with GPT2, we conducted objective and subjective evaluations by comparing the following methods: (1)~\textbf{\textit{Ground-truth}}, ground-truth audio clips included in the test data; (2)~\textbf{\textit{Full-sentence}}, sentence-level Tacotron2 model~\cite{shen17tacotron2}; (3)~\textbf{\textit{Independent}}, where the TTS model synthesized a current speech segment independently of the contextual information~\cite{Yanagita19itts}; (4)~\textbf{\textit{Unicontext}}, where the TTS model used only the past observed segment for context conditioning of the TTS model; (5)~\textbf{\textit{Bicontext}}, which is the proposed method described in Section~\ref{methods} without the fine-tuning method; (6)~\textbf{\textit{Bicontext (truth)}}, where we used ground-truth test transcripts for unobserved future sentences like the conventional lookahead-$k$ strategy~\cite{Ma20prefix} that waits for observing $k$ words; and (7)~\textbf{\textit{Bicontext (fine-tuned)}}, which applied the fine-tuning method to \textit{Bicontext}.
Audio samples\footnote{\url{https://takaaki-saeki.github.io/itts_lm_demo/}} synthesized with these methods are publicly available.

\vspace{-2mm}
\subsection{Objective evaluations}
Unlike the utterance-level TTS, incremental TTS is more prone to fail in synthesis and to output non-recognizable speech.
Therefore, we measured the word error rate (WER) and character error rate (CER), defined as the word- and character-level Levenshtein distance~\cite{Levenshtein_SPD66}, using the state-of-the-art ASR model to evaluate how natural and easy the output speech is to recognize as a human utterance.
We used a joint-CTC Transformer-based model~\cite{kim2018jointctc} trained on librispeech~\cite{librispeech}, which is included in ESPnet~\cite{espnet}.
Table~\ref{tab:result} lists the results.

First, both the CER and WER were vast for \textit{Independent}. 
In some cases, the \textit{Independent} did not predict the stop flag correctly due to the lack of contextual information, which caused a sluggish part in the output speech and significantly increased the insertion rate.
As a result, \textit{Bicontext} synthesized output speech that was easier to recognize than that with \textit{Independent}. 
Furthermore, the error rates of \textit{Bicontext} were lower than those of \textit{Unicontext}, which used only the observed context, thus demonstrating the effectiveness of the pseudo lookahead with GPT2 for incremental TTS.
Finally, examining \textit{Bicontext (fine-tuned)}, we can see that the fine-tuning method decreased the error rates to a level comparable to that of \textit{Bicontext (truth)}, which used the test transcript for the lookahead.

\begin{table}[t]
\centering
\caption{CER, WER, and MOS for each method described in Section~\ref{sec:evaluation_cases}.}
\vspace{-1mm}
\label{tab:result}
\scalebox{1.1}{
\begin{tabular}{l|ccc}
\hline
Method                & CER     & WER  & MOS     \\ \hline
\textit{Ground-truth}            & 5.1 \%  & 17.9 \% & $4.28 \pm 0.13$\\
\textit{Full-sentence}          & 5.5 \%  & 18.2 \% & $3.82 \pm 0.12$  \\
\textit{Bicontext~(truth)} & 8.2 \%  & 24.2 \% &  $3.36 \pm 0.16$ \\ \hline
\textit{Independent}            & 38.9 \% & 96.9 \% & $2.69 \pm 0.20$ \\
\textit{Unicontext}             & 22.8 \% & 53.9 \% & $2.99 \pm 0.18$ \\
\textit{Bicontext}              & \textbf{11.9} \% & \textbf{29.8} \% & $3.38 \pm 0.14$ \\ 
\textit{Bicontext (fine-tuned)} & \textbf{8.0} \% & \textbf{22.5} \% & $3.44 \pm 0.16$ \\ \hline
\end{tabular}
}
\vspace{-8pt}
\end{table}
\vspace{-2mm}
\subsection{Subjective evaluations}
To evaluate the quality of output speech, we conducted a mean opinion score (MOS) evaluation test~\cite{steijl16mos} on naturalness.
Forty listeners recruited through Amazon Mechanical Turk~\cite{mturk} participated in the evaluation, and each listener evaluated 35~speech samples, where we randomly chose five samples from the output utterances of the test data for each method.
Table~\ref{tab:result} shows the average MOS scores with 95~\% confidence intervals.

First, our proposed methods scored significantly higher than \textit{Independent}, which is based on the prior work~\cite{Yanagita19itts}. 
Furthermore, the proposed methods outperformed \textit{Unicontext}, which considered only the past observed context, thus demonstrating that the pseudo lookahead with GPT2 significantly improves the naturalness of synthesized speech.
When we compare the proposed methods, \textit{Bicontext} and \textit{Bicontext (fine-tuned)}, the average score of \textit{Bicontext (fine-tuned)} was higher, suggesting that language model-guided fine-tuning leads to more effective pseudo lookahead generation.
Finally, our proposed methods achieved naturalness comparable to \textit{Bicontext (truth)}, which uses the lookahead information (like the method of Ma et al.~\cite{Ma20prefix}).
This result indicates that the pseudo-lookahead conditioning with a language model-guided fine-tuning improves the quality equivalently to waiting for the actual lookahead observations without increasing the latency.
Note that we also conducted AB tests for \textit{Bicontext}, \textit{Bicontext (fine-tuned)}, and \textit{Bicontext (truth)}, but we did not find any significant differences between them as in the MOS evaluation test.

\vspace{-2mm}
\section{Conclusion}

In this letter, we proposed an incremental text-to-speech (TTS) method using the pseudo lookahead generated with a large pretrained language model.
This method synthesizes a current speech segment while generating the unobserved future information with GPT2 instead of waiting for its actual observation.
We also proposed a language model-guided fine-tuning method to use the pseudo lookahead for incremental TTS more effectively.
Experimental results demonstrated the effectiveness of our methods in terms of both the synthetic speech quality and the latency.
In future work, we will further enhance our method towards an incremental TTS with a quality equivalent to sentence-level TTS using only observed information.

\bibliographystyle{IEEEbib}
\bibliography{tts}

\end{document}